\documentclass[aps,prb,showpacs,twocolumn,superscriptaddress]{revtex4}

\usepackage{graphicx}
\usepackage{amsmath}
\usepackage{natbib}
\usepackage{bm}
\usepackage[dvips]{color}

\begin{document}

\title{Rientrant metallicity in  the Hubbard model: the case of  honeycomb nanoribbons}
\date{\today}

\author{F.~Manghi}
\email[Electronic address: ]{franca.manghi@unimore.it}
\affiliation{Dipartimento di Fisica, Universit\`a di Modena e Reggio Emilia, Via Campi 213/A, I-41125 Modena, Italy }
\affiliation{CNR - Institute of NanoSciences - S3}

\author{F. Petocchi}
\affiliation{Dipartimento di Fisica, Universit\`a di Modena e Reggio Emilia, Via Campi 213/A, I-41125 Modena, Italy }
\begin{abstract}

Based on the Cluster Perturbation solution of the Hubbard hamiltonian for a 2-D honeycomb lattice we present quasiparticle band structures of nanoribbons at half filling as a function of the on-site electron-electron repulsion. We show that at moderate values of e-e interaction  ribbons with armachair shaped edges exhibit an unexpected semimetallic behavior, recovering the original insulating character only at  larger values of $U$.

\end{abstract}
 \maketitle

The repulsive interaction among electrons is responsible of the failure of single particle picture and of the opening/widening of energy gaps in solids. The Hubbard model is the  paradigm to describe this phenomenon: sufficiently large values of the on-site e-e  repulsion  inhibit the inter-site hopping  favoring in this way an insulating behaviour.
The 1-atom thick  2D  honeycomb lattice (graphene) does not contradict this picture:  many body effects due to on-site Coulomb repulsion have been  shown to lead, for sufficiently strong interactions, to semimetal-to-insulator transition \cite{Liebsch,WeiWu} as well as to other deviations from Fermi-liquid behavior such as unconventional quasiparticle lifetimes \cite{Guinea}, long range antiferromagnetic order \cite{Paiva} and spin liquid phase \cite{Meng}.

In this paper we show that for  honeycomb nanoribbons the repulsive e-e interaction may be   responsible  of a metallic phase  in  ribbons that in the single particle picture are semiconducting. This appears to be another extraordinary property of the 2-D honeycomb lattice.

It is well known that  honeycomb nanoribbons manifest peculiar properties related to the topology of their edges \cite{RevModPhys.81.109}: according to single-particle theory  ribbons with armchair shaped edges may exhibit a finite energy gap depending on their width \cite{edgestates,Ezawa,Fertig}, while ribbons with zigzag  edges  are metallic and become insulating only after the inclusion of an antiferromagnetic order \cite{Mcohen,Pisani}. The modifications of the single particle band structure of zigzag graphene ribbons due to e-e interaction has been investigated within a mean field solution of the Hubbard model  \cite{Rossier,Yamashiro}, showing spin polarization of edge states and gap opening at the Fermi level. How the single particle band picture evolves to a quasi-particle one and how the electronic states of both armchair and zigzag honeycomb ribbons are modified by  on-site Coulomb repulsion  described as a true many body term is the question we address in this paper.

We have adopted a  many body approach based on the Cluster Perturbation Theory  \cite{Senechal} (CPT). CPT belongs to the class of Quantum Cluster theories\cite{RevModPhysQC} that solve the problem of many interacting electrons in an extended lattice by a  \emph{divide-and-conquer } strategy, namely solving first the many body problem in a subsystem of finite size and then embedding it within  the infinite medium. Quantum Cluster theories represent some of the most powerful tools for the numerical investigation of strongly correlated many-body
systems. They include Dynamical Cluster Approach \cite{DCA}, Cellular Dynamical Mean Field Theory  \cite{Senechal_CDMFT} as well as CPT and have found an unified language within the
variational scheme  \cite{VCA} based on the the Self Energy Functional approach \cite{SFA}. CPT  has many interesting characteristics  and  gives access to non trivial many body effects in a relatively simple way: it exactly reproduces the limits $U/t=0$ (non-interacting band limit), $ U/t =\infty$ (atomic limit) and  the Mott-Hubbard metal-to-insulator transition for intermediate values of $U/t$; it recovers most of the characteristics of  the exact solution of the 1-dimensional case\cite{Senechal,VCA_1D}; finally it is relatively easy to implement  and, at least for simplest systems, without much numerical effort. In its variational form \cite{VCA,Balzer} (Variational Cluster Approximation (VCA) ) it can be applied to  systems with spontaneously broken symmetry  describing antiferromagnetism  \cite{VCA2} and superconductivity \cite{Aichhorn,Senechal_VCA}. In zigzag  honeycomb ribbons VCA has been applied to study the transition from  topological to antiferromagnetic insulator \cite{Yu_ribbon_TI}.

\begin{figure}[htbp]
\includegraphics[width=5cm]{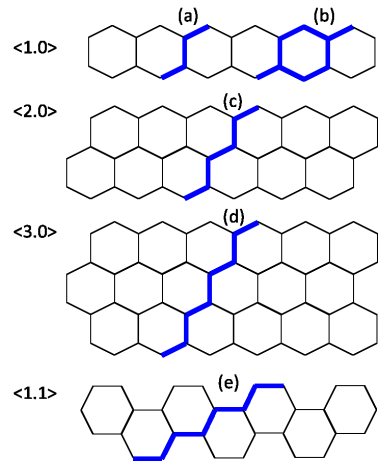}
  \caption{\label{geometry} (Color online) Geometric configurations of  honeycomb ribbons with zigzag and armchair edges identified according to the notation  of   reference \cite{Ezawa}. The single (a,c,d,e) and double (b) chains used to \emph{tile} the extended 1D lattice are also shown.}
\end{figure}

In CPT \cite{Senechal} the lattice is  seen as the periodic repetition of identical clusters (Fig. \ref{geometry} ) and the Hubbard Hamiltonian is written as the sum of two terms, an intra-cluster ($ \hat{H_c}$) and an inter-cluster one ($\hat{V} $)
\begin{equation*}
\hat{H} = \sum_{l} \hat{H}_{l}+ \sum_{l\neq l'} \hat{H}_{l l'}  = \hat{H_c}+\hat{V}
\end{equation*}
where  the summations are over all the clusters and
\begin{eqnarray*}
\nonumber
\hat{H}_l &=& -t\sum_{i j \sigma} \hat{c}_{i l \sigma}^{\dag} \hat{c}_{j l \sigma} +
U \sum_{i } \hat{c}_{i l \uparrow}^{\dag}\hat{c}_{i l \uparrow}\hat{c}_{i l \downarrow}^{ \dag}\hat{c}_{i l \downarrow} \\ \nonumber
\hat{H}_{l l'} &=& -t\sum_{i j \sigma} \hat{c}_{i l\sigma}^{\dag}\hat{c}_{j l' \sigma}
\end{eqnarray*}
Since in the Hubbard model the e-e Coulomb interaction is on-site, the inter-cluster hamiltonian $\hat{V} $ is single particle and the  many body term  is present in the intra-cluster hamiltonian $ \hat{H}_{c}$ only, a key feature for the practical implementation of the method.
Having partitioned the Hamiltonian in this way we may write the resolvent operator $ \hat{G} $ as
$\hat{G}^{-1}=  z-\hat{H_c}-\hat{V} = \hat{G^c}^{-1}-\hat{V}$ and from this
 \begin{equation}\label{Dyson}
\hat{G} =   \hat{G^c} + \hat{G^c} \hat{V} \hat{G}
 \end{equation}
The one-particle propagator
\begin{eqnarray*}
 \mathcal{G}(\textbf{k} n  \omega)&=& <\Psi_0|\hat{c}_{\textbf{k}  n }^{\dag}\hat{G} \hat{c}_{\textbf{k} n }|\Psi_0> \\ \nonumber
& +& <\Psi_0|\hat{c}_{\textbf{k}  n }\hat{G} \hat{c}_{\textbf{k} n }^{\dag}|\Psi_0>
\end{eqnarray*}
is obtained exploiting the transformation from localized to Bloch basis
\begin{equation*}
\hat{c}_{\textbf{k}  n}^{\dag}=\frac{1}{\sqrt{N}}\sum_{i l}\alpha^n_i(\textbf{k})^* e^{-i\textbf{k}\cdot(\textbf{R}_l+\tau_i)}\hat{c}_{i l }^{\dag}
\end{equation*}
and similarly for $\hat{c}_{\textbf{k}  n  } $. Here   $\alpha^n_i(\textbf{k})$ are the eigenstate coefficients obtained by a band calculation for a superlattice of $L$ identical clusters, identified by the lattice vectors $R_l$, each cluster containing M sites  at positions $\tau_i$; $n$ is the band index  and the summation is over the $N=L\times M$ lattice sites. We get
\begin{equation}\label{gk}
 \mathcal{G}(\textbf{k} n  \omega)=
  \frac{1}{M}\sum_{i i'} e^{-i\textbf{k}\cdot(\tau_i-\tau_{i'})}|\alpha^n_i(\textbf{k})|^2 \mathcal{G}_{i i'}(\textbf{k}   \omega)
\end{equation}
where $\mathcal{G}_{i i'}(\textbf{k}   \omega)$ is the superlattice Green function, namely the Fourier transform of the Green function in local basis
\begin{equation*}
\mathcal{G}_{i i'}(\textbf{k}   \omega)=\frac{1}{L}\sum_{l l'}e^{-i\textbf{k}\cdot(\textbf{R}_l-\textbf{R}_{l'})}
\mathcal{G}^{l l'}_{i i'}(  \omega)
\end{equation*}
This is the quantity that can be calculated  by eq.\ref{Dyson}
\begin{equation}\label{QC}
\mathcal{G}_{i i'}(\textbf{k}    \omega)=\mathcal{G}^c_{ i i'}(  \omega)+ \sum_j B_{i j}(\textbf{k}  \omega) \mathcal{G}_{j i'}(\textbf{k}   \omega)
\end{equation}
where  $M\times M$ matrix $B_{i j}(\textbf{k}   \omega)$
is the Fourier transform of $\hat{G}^c\hat{V}$ involving neighboring sites that belong to different clusters.
Once the cluster Green function in the local basis  $\mathcal{G}^c_{ i i'}(  \omega)$ has been obtained    by exact diagonalization, eq. \ref{QC} is solved quite simply  by a $M\times M$  matrix inversion at each $\textbf{k}$ and $\omega$. The quasi particle spectrum is then obtained in terms of  spectral function $A(\textbf{k}  \omega)$
\begin{equation*}
    A(\textbf{k}   \omega) = \frac{1}{\pi}\sum_n Im  \mathcal{G}(\textbf{k} n \omega).
\end{equation*}
The key approximation in this derivation has been  to identify the  many electron ground states of the extended lattice as the product of cluster few electron ones.
This is certainly wrong except at $U/t=0 $ or $U/t= \infty$ and at intermediate values of $U/t$ it is important to verify the accuracy of the results by using larger and larger cluster sizes. In practice this procedure is limited by the dimensions of Hilbert space  used in the exact diagonalization, dimensions that grow exponentially with the number of  sites.

\begin{figure}[htbp]
\includegraphics[width=7cm]{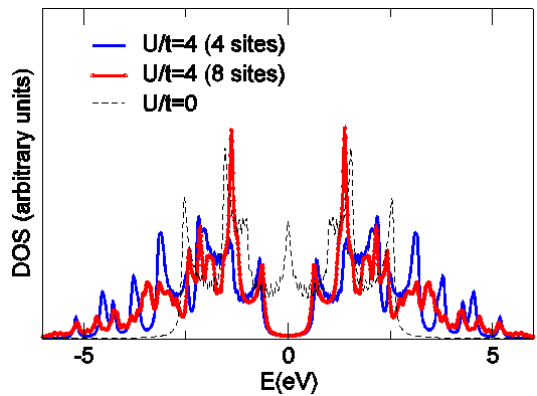}
  \caption{\label{dos} (Color online) Density of Quasi Particle states obtained for $U/t=4$ for a zigzag terminated ribbon assuming different cluster sizes (Fig  \ref{geometry} a,b) compared with non-interacting DOS. }
\end{figure}

Fig. \ref{geometry} illustrates the  ribbon geometries and the  open chains that are periodically repeated  to reproduce both zigzag and armchair ribbons.
We have considered  zigzag ribbons of different widths and in the case of ribbon $<1.1>$  both single chain (4 sites) and double chain (8 sites) tilings in order to test the influence of cluster size on the quasiparticle spectrum. In  Fig. \ref{dos} we show the  QP Density Of States (DOS) obtained for a specific value of $U/t$  and the two above mentioned cluster sizes, compared with the corresponding non-interacting DOS. We notice that the main features (peak positions, gap opening)  do not depend much on the cluster size and we may confidently use the smallest cluster size in all cases.

\begin{figure}
\includegraphics[width=7cm]{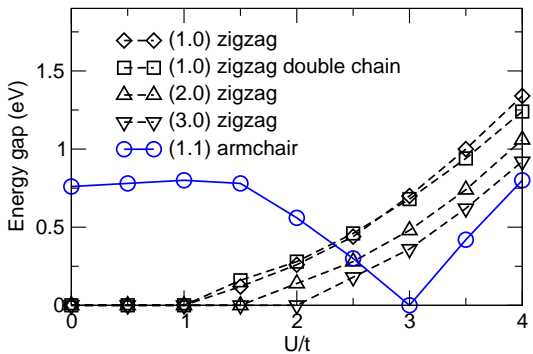}
  \caption{\label{gap} (Color online) Value of the energy gap as a function of $U/t$ showing a drastically  different behavior in zigzag and armchair  ribbons.  For zigzag termination results obtained for different cluster sizes (single and double chain, Fig. \ref{geometry} (a) and (b) ) and for different ribbon widths are shown. }
\end{figure}

As a result of  the inclusion of e-e interaction extra structures  appear in the quasi particle spectrum  below (above) filled (empty); this is  what happens also in real materials where Hubbard correlation may be responsible of severe  energy renormalization,   quasiparticle quenching \cite{manghiCo,manghifink} and the appearance of short lived satellites structures \cite{manghiNi}.  The main effect of Hubbard correlation however is the opening, for sufficiently strong interaction $U \geq U_c$, of a well defined  gap. This is due to the well known  Mott-Hubbard mechanism: $U$ inhibits double occupancies of sites  and in this way makes  electron hopping from site to site less and less energetically favorable, driving the system  across a metal-to-insulator transition. This is what happens in  model systems at half occupation and in real materials  \cite{RevModPhys.70.1039}.
As shown in Fig. \ref{gap} the value of  $U_c $ for zigzag ribbons depends on the ribbon width.  Previous VCA calculations \cite{Yu_ribbon_TI} have found for wider ribbons  $U_c \approx 3t$, a results not too far from the present one taking into account  the larger ribbon width and  the variational procedure used in that work. Recent ab-initio estimate  of the screened on-site Coulomb interaction in graphene \cite{Wehling} provide values of the the same order ($U/t\sim 3.5$).

\begin{figure}[htbp]
\includegraphics[width=7cm]{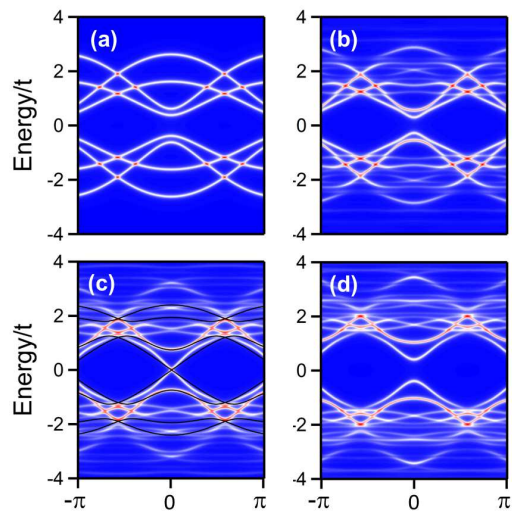}
  \caption{\label{aknarmch} (Color online) Quasiparticle band structure for armchair terminated ribbon $<$1.1$>$ obtained with  $U/t=0$ (a),$= 2$ (b),$= 3$ (c), $=4 $(d). The black  line superimposed in (c) is the band structure obtained in the single particle picture assuming different hopping terms (see text). }
\end{figure}

For the armchair terminated ribbon the situation is completely different. As previously mentioned, according to single particle theory  ribbons with armchair shaped edges may be metallic or insulating depending on their width \cite{edgestates,Ezawa,Fertig}.  In particular the   ribbon of interest here   exhibits a finite gap in the non-interacting picture and we may  expect the one-site  e-e interaction to reinforce this insulating behavior enlarging the gap.
This, surprisingly enough, is not quite the case: switching on e-e correlation  the gap first diminishes, reaches zero at $U/t =3$ and only after that  grows linearly with $U/t$ (Fig. \ref{gap} ). At $ U=U_c$ the QP band dispersion (Fig. \ref{aknarmch}) becomes linear  around $\textbf{k}=0$ and the system semimetallic.

We may look at the character of quasiparticle states by considering the local spectral function, namely $A(\textbf{k},i,\omega)=
\frac{1}{\pi}\sum_n Im  \mathcal{G}(\textbf{k} n \omega)\mid \alpha^n_i(\textbf{k})\mid ^2$. We notice that the effect of e-e repulsion is to delocalize states close to the gap region that for  $U/t=0$  are localized either at the edges  or in the
inner part of the ribbon. In particular for $U/t=3$ the localization of
gapless states responsible of the semimetallic behavior is equally distributed  across the ribbon.
We have found that it is  possible to reproduce the same gapless band dispersion with the same delocalized character in  a single particle picture attributing different hopping terms to edge sites and to inner ones:
this is shown in Fig. \ref{aknarmch}(c)  assuming  hopping between edge states  exactly twice as big as  hopping between inner ones.  In this sense the net effect  of the on-site e-e repulsion  is  to renormalize the inter-site hopping selectively across the ribbon and to make hopping between doubly coordinated sites (edge sites) more favorable than hopping between
sites with triple coordination (inner sites). The gap closing would then be due to a mechanism similar to a mechanical strain. The analogy between mechanical strain, e-e repulsion and magnetic field as responsible of gap tuning in graphene has been recently investigated  \cite{geim,PhysRevLett.107.106402}  showing in particular that in the mean field approximation a local nearest neighbor Coulomb interaction may create in a 2D honeycomb lattice non trivial magnetic configurations and  metallic phases with broken time reversal symmetry \cite{PhysRevLett.107.106402}. In the present study we have shown that also on-site Coulomb repulsion if treated as a full many body term may induce a semimetallic behavior in a semiconducting 1D honeycomb lattice.

\begin{figure}
\includegraphics[width=7cm]{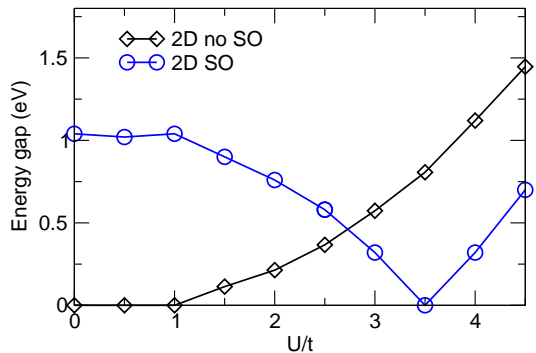}
  \caption{\label{gap_SO} (Color online) Energy gap as a function of $U/t$ for 2D honeycomb lattice with (circles) and without spin-orbit (diamonds) interaction. }
\end{figure}

What is surprising is that the Hubbard interaction has  opposite effects in the two types of honeycomb ribbons: it \emph{opens}  a gap in the metallic ribbon and \emph{closes} it in the semiconducting one.
A  similar behavior has been reported in a recent paper on correlation effects in topological insulators \cite{Yu_ribbon_TI}. Starting from an extended 2D honeycomb lattice made semiconducting by spin-orbit (SO) interaction the authors show that for $U/t=3$ the existing gap closes down. We have repeated the calculation using the same SO parameter and calculated also in this case the  gap as a function of $U/t$. The results are shown in Fig. \ref{gap_SO} for a 2D honeycomb lattice with and without SO and again in the semiconducting system we find  a regime  of $U/t$ where  the energy separation between filled and empty states decreases and the pre-existing gap closes down to zero. And this even if  the original semiconducting behavior has a completely different physical origin (SO interaction in  2D  instead of armchair termination in 1D).

In conclusion, we have found that many body effects associated to local e-e repulsion may be responsible of a (semi)metallic phase in systems with honeycomb lattice. The Hubbard mechanism  that inhibits double occupancies of sites,  instead of reducing the ability of  electrons to jump from site to site, induces a selective renormalization of inter-site hopping and the energy separation between filled and empty states becomes zero at a specific k-point. This behavior appears to be a characteristic of honeycomb topology - or perhaps more generally of bipartite lattices - at half occupation  and suggests the existence of an $U$-dependent additional symmetry  \cite{lieb}. The Electrons trapped in artificial  lattices with honeycomb  geometry \cite{gibertini} are the best candidates where this anomalous behavior induced by e-e repulsion (gap closing/opening for semiconducting/metallic systems) can be experimentally verified.

\end{document}